\documentclass[12pt]{iopart}

\usepackage{graphicx}
\usepackage[usenames]{color}
\newcommand{\beq}{\begin{equation}}
\newcommand{\eeq}{\end{equation}}
\newcommand{\beqa}{\begin{eqnarray}}
\newcommand{\eeqa}{\end{eqnarray}}
\newcommand{\ba}{\begin{array}}
\newcommand{\ea}{\end{array}}

\begin{document}

\title[Two-dimensional dipolar Bose-Einstein condensate bright and
vortex solitons]{Two-dimensional dipolar Bose-Einstein condensate bright and 
vortex solitons on one-dimensional optical lattice}


\author{S. K. Adhikari$^1$\footnote{Email: adhikari@ift.unesp.br; URL:
http://www.ift.unesp.br/users/adhikari/}, and P. Muruganandam$^{1,2}$\footnote{anand@cnld.bdu.ac.in}}
\address{$^1$Instituto de F\'{\i}sica Te\'orica, 
UNESP - Universidade Estadual 
Paulista, 01.140-070 S\~ao Paulo, S\~ao Paulo, Brazil}
\address{$^2$School of Physics, Bharathidasan University,
Palkalaiperur Campus,
Tiruchirappalli 620024,
Tamilnadu,
India}

\begin{abstract} 

By solving
the 
three-dimensional Gross-Pitaevskii equation 
we generate  two-dimensional   axially-symmetric and anisotropic dipolar
Bose-Einstein 
condensate bright   solitons, for repulsive atomic interaction, 
stabilized 
by only 
a weak one-dimensional optical lattice (OL) 
aligned along and  perpendicular, respectively,  to 
the dipole polarization direction. 
In the former case 
vortex   solitons can also be created.
We show that it is possible to make 
a stable  array of small  interacting axially-symmetric dipolar
solitons put on alternate OL sites. Further, we demonstrate 
the elastic nature of the collision of two such solitons.

 \end{abstract}


\maketitle

A bright soliton is a self-reinforcing solitary wave  
that maintains its shape, while traveling at constant speed, due 
to a cancellation of nonlinear attraction and dispersive effects. {
Integrable} solitons without any external trap or intervention { for cubic nonlinearity}
exist only in 
one dimension (1D).  
Experimentally, bright matter-wave solitons and soliton trains were 
created in a quasi-1D
 Bose-Einstein condensate (BEC) of $^7$Li \cite{4r,4rb} 
and $^{85}$Rb atoms \cite{5r} by turning the atomic interaction attractive from 
repulsive using  a Feshbach 
resonance (FR) \cite{fesh} and employing a transverse trap. 

Although, the normal three-dimensional (3D)  BEC soliton \cite{perez}   is of great interest, 
such a BEC has only short-range 
attraction which makes it vulnerable against collapse. 
 Physical systems are stable due to a peculiar nature of interaction among its constituents
(atoms, molecules and nuclei), e.g.,
short-range repulsion and long-range attraction. 
Lately,  BEC of $^{52}$Cr \cite{pfau,rpp} and $^{164}$Dy \cite{dyspro,dy1}
atoms with a large long-range
dipolar interaction has been observed. Also, experimental tuning of the long-range 
dipolar interaction  by means of rapidly 
rotating orienting fields \cite{condip} as well as 
of the short-range  atomic interaction using a FR \cite{fesh}
are completely under control.
This engineering of the atomic and dipolar interactions makes the dipolar BEC (DBEC) 
an interesting system
for the formation of   soliton \cite{Tikhonenkov2008,adhisol,ps}. 
The long-range anisotropic dipolar interaction is attractive in some 
directions and repulsive in others. If it were attractive in all directions, stable  
robust 3D
DBEC 
solitons, corresponding to a minimum in energy functional,
would naturally be formed  for repulsive short-range atomic and 
attractive long-range dipolar interactions \cite{rydberg}. 


Normal  dipolar interaction leads to attraction along the polarization $z$ direction
and repulsion along 
transverse directions. It is possible to have the opposite by tuning the dipole interaction to 
``negative" values by orienting fields \cite{condip} 
and this set-up was used in some studies \cite{ps,Pedri2005}.   
 For normal dipolar interaction,
an anisotropic two-dimensional  (2D) soliton can be obtained for 
repulsive { short-range} atomic interaction
if a weak OL is placed along  $y$ axis,
perpendicular to the polarization direction $z$, 
to overcome the  dipolar repulsion in transverse directions.   For dipolar
interaction tuned to negative values \cite{condip},  axially-symmetric 2D bright and vortex solitons can be obtained for repulsive 
{ short-range}
atomic interaction
if a weak OL is placed along  $z$ 
axis
to overcome the   dipolar repulsion in that direction in this sign-changed setting.
In all cases, the 
dipolar repulsion is weak and we do not need any trap in other directions to stabilize a   
soliton. Such 2D solitons cannot be stabilized without the dipolar interaction \cite{ps}.

We  present a linear stability analysis for the 
axially-symmetric soliton 
  \cite{stability}.
 We 
study the  2D DBEC solitons
 using the numerical and Lagrangian variational analysis  
of the 3D Gross-Pitaevskii (GP) equation.  
The effective Lagrangian of the variational analysis has the same 
structure as that of a  
generalized classical dynamical   system with two degrees of freedom. 
We find that stable  (referred to as  ``center" as it corresponds to a stable
periodic orbit around a center in phase space) and {unstable stationary}
(called ``saddle" as it corresponds to a saddle point in   energy) states 
appear and disappear  through the   mechanism of  { \it saddle-center
bifurcation}   \cite{scb}.

There have been studies of 2D DBEC solitons with strong harmonic traps along $y$  \cite{Tikhonenkov2008} or 
 $z$ \cite{ps}
axis
 and of 1D DBEC solitons under transverse harmonic trap \cite{adhisol}. The present solitons confined 
by only a weak OL along $y$ or $z$ axis, respectively, are distinct.
The previous studies \cite{Tikhonenkov2008,ps,Pedri2005}
will essentially have {an approximate}   Gaussian density distribution along the infinite trap direction, whereas
the present solitons 
will have an exponential density distribution due to weak finite traps in these directions.   
More interestingly,  an OL simulates the periodic electron-atom potential in a solid and the study of solitons 
in an OL  is also of interest in condensed-matter physics \cite{band,lewen}.
We show that a new type of  stable interacting 1D
array of solitons can be 
formed in 3D space when  tiny  axially-symmetric  interacting DBEC bright 
solitons are placed on alternate sites of the OL.
However, if the solitons are placed
on all sites of the OL, the array is destroyed due to strong long-range dipolar interaction among its 
constituents.   
{Statics and dynamics of
such periodic array of tiny droplets of dipolar matter 
are of concern in condensed matter physics \cite{lewen}, as 
they simulate many problems of general interest, such as, a periodic linear array of tiny magnets. Polarized 
droplets of $^{52}$Cr and $^{164}$Dy have permanent magnetic dipole moment.}
 
In a repulsive BEC on 3D OL, gap solitons having negative effective mass
responsible for attraction,
with the chemical potential lying in the band-gap, can be made \cite{gs,bier}. The present solitons on 1D OL, 
free to move in 
the transverse plane  are bright, and not gap, solitons.

We consider  a   
 DBEC  of $N$ atoms, each of mass $m$, using the  GP
equation: \cite{pfau}
\begin{eqnarray}  \label{gp3d} 
i \frac{\partial \phi({\bf r},t)}{\partial t}
& =&  \left[ -\frac{\nabla^2}{2} + V_{\mbox{OL}}^{1D} +g|\phi|^2+
F\right] \phi({\bf r},t)
\end{eqnarray} 
with $g=4\pi a N$, $ F=\int U_{dd}({\bf r -r'})|\phi({\bf r'},t)|^2d
{\bf r'}$,  ${\bf r}\equiv \{x,y,z\} \equiv \{\rho,z \}$,
$V_{\mbox{OL}}^{1D}\equiv -V_z\cos(2z)$ or $-V_y\cos(2y)$ is the weak OL for stabilizing the 
soliton, 
$ U_{dd}({\bf R}) =  
g_{dd}(1-3\cos^2\theta)/R^3, g_{dd}=3a_{dd}N\alpha $,
 ${\bf R=r-r'},$  
 normalization $\int \phi({\bf r})^2 d {\bf r}$ = 1,
   $a$ the  scattering length, $\theta$ 
the angle between $\bf R$ and   $z$,  $a_{dd}
=\mu_0\bar \mu^2 m /(12\pi \hbar^2)$ 
the strength of 
dipolar interaction,  
 $\bar \mu$ the (magnetic) dipole moment of an   atom, and $\mu_0$ 
the permeability of free space. The parameter $\alpha$ $ (1>\alpha>-1/2) $ can be tuned by a rapidly   rotating magnetic field 
 allowing the change of the sign of   dipole interaction.
  In   (\ref{gp3d}), 
length is measured in 
units of  $l_0 \equiv  \lambda/(2\pi)$, time $t$ in units of $t_0 = ml_0^2/\hbar$, $V_y,V_z,$ and energy  in units of $2E_R$, where 
$E_R=h^2/(2m\lambda^2)$ is recoil energy,
with $\lambda$  the OL wave length.



First we consider the axially-symmetric soliton for 
$ V_{\mbox{OL}}^{1D}=  -V_z\cos(2z)$ and $ \alpha<0$. In this case the  
Lagrangian density of 
  (\ref{gp3d})    is \cite{jb,vortex}
\begin{eqnarray}
{\cal L}=& \,\frac{1}{2}i\left( \phi \phi^{\star}_t
- \phi^{\star}\phi_t \right) +\frac{1}{2}\vert\nabla\phi\vert^2
+   2\pi aN\vert\phi\vert^4+V^{1D}_{\mbox{OL}}|\phi|^2
\nonumber \\ & \,
+ \frac{1}{2}N\vert
\phi\vert^2\int U_{dd}({\mathbf r}-
{\mathbf r'})\vert\phi({\mathbf r'})\vert^2 d{\mathbf r}'
.\label{eqn:vari}
\end{eqnarray}
For a variational study we
use the Gaussian ansatz  \cite{jb,vortex}: 
$ \phi({\bf
r},t)=
\exp(-
{\rho^2}/{2w_\rho^2}- {z^2}/{2w_z^2}$ $ +i\gamma\rho^2
+i\beta z^2 )  /({w_\rho \sqrt w_z}\pi^{3/4})$
where $w_\rho$ and $w_z$ are time-dependent widths and 
$\gamma$ and $\beta$ are time-dependent {chirps}. 
The
effective Lagrangian $L$ (per particle) is
\begin{eqnarray}\label{lag}
L &\equiv  &  \int {\cal L}\,d{\mathbf r}
 =  \left(w_\rho^2\dot{\gamma} +\frac{1}{2}
w_z^2\dot{\beta}+2w_\rho^2\gamma^2+w_z^2\beta^2          \right) 
\nonumber \\ 
&+&  E_{\mathrm{kin}}+ E_{\mathrm{trap}}+  E_{\mathrm{int}}
 , 
\end{eqnarray} 
with kinetic, trap, and interaction energies given, respectively,  by  
$  E_{\mathrm{kin}}=  
({1}/{2w_\rho^2} + {1}/{4w_z^2}), 
E_{\mathrm{trap}}= -V_z\exp(-w_z^2), $ $
  { E}_{\mathrm{int}}= N[a-a_{dd}f(\kappa)]/(\sqrt{2 \pi}w_\rho^2w_z)  , $
where 
$  
f(\kappa)=[1+
2\kappa^2-3\kappa^2$ $d(\kappa)] /
(1-\kappa^2), d(\kappa)
=(\mbox{atanh}\sqrt{1-\kappa^2})/\sqrt{1-\kappa^2}, \kappa=w_\rho/w_z.$
The Euler-Lagrange equations for parameters 
$ w_\rho, w_z, \gamma, \beta$
can be used to obtain the following equations of
the widths for the dynamics of the DBEC state 
\begin{eqnarray} \label{f3}  &&
\ddot{w}_{\rho} =
\frac{{{1}}}{w_\rho^3} +\frac{
1}{\sqrt{2\pi}} \frac{N}{w_\rho^3w_{z}}
\left[2{a} - a_{dd}{e(\kappa) }\right],
\label{eq:dimless:a} \\ && \ddot{w}_{z} =
\frac{1}{w_z^3}+ \frac{ 1}{\sqrt{2\pi}}
\frac{2N}{w_\rho^2 w_z^2} \left[{a}-
a_{dd}h(\kappa)\right]  -\frac{4V_zw_z}{\exp(w_z^2)} , \label{f4} \end{eqnarray}
with $e(\kappa)=[2-7\kappa^2-4\kappa^4+9\kappa^4
d(\kappa)]/(1-\kappa^2)^2, h(\kappa)= [1+10\kappa^2-2\kappa^4-9\kappa^2
d(\kappa)]/(1-\kappa^2)^2.$
The widths of a stationary  soliton 
of energy  $E\equiv E_{\mathrm{kin}}+ E_{\mathrm{trap}}+ E_{\mathrm{int}}$
are obtained by solving 
 (\ref{f3}) and (\ref{f4}) for $\ddot w_\rho=\ddot w_z=0$.

\begin{figure}
\begin{center}
\includegraphics[width=\linewidth,clip]{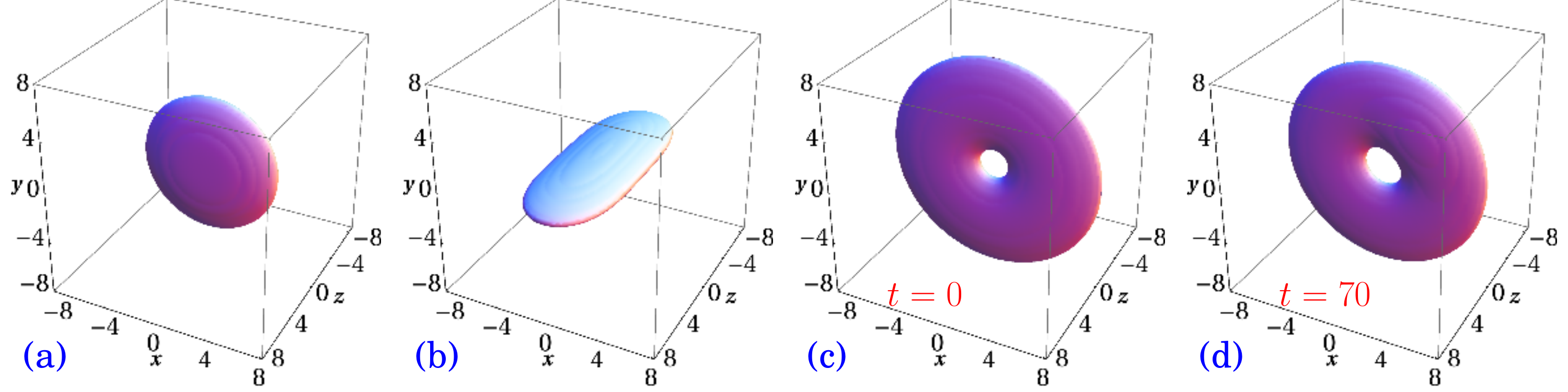}
\end{center}

\caption{(Color online) (a) Numerical  3D contour of an axially-symmetric
soliton for $g=50$ and $g_{dd}=-15$ on OL 
$V^{1D}_{\mbox{OL}}= -2\cos(2z)$. (b) The same for  an anisotropic soliton 
with $g=50$ and $g_{dd}=20$ on OL  $V^{1D}_{\mbox{OL}}=-2\cos(2y)$ and  
that  for a vortex soliton with $g=5$ and $g_{dd}=-9$ on OL
$V^{1D}_{\mbox{OL}}=-2\cos(2z)$ at $t=$ (c) $0$ and (d) $ 70$.
 The density 
$|\phi({\bf r})|^2$ on the 
contour is  0.001.
}
\label{fig1}
\end{figure}

To obtain a quantized vortex of unit angular momentum $\hbar$; 
around $z$ axis,  we
introduce a phase (equal to the azimuthal angle) in 
wave function \cite{vortex2}. This
procedure introduces a centrifugal term $1/[2(x^2+y^2)]$ in the 
GP equation for a vortex and we
adopt this method to study an axially-symmetric vortex soliton
on a 1D OL along $z$ axis for the dipolar interaction tuned to 
negative values.

For the  anisotropic 2D soliton on OL  
$ V^{1D}_{\mbox{OL}}= -V_y\cos(2y)$ with $\alpha >0$, we consider 
 a minimization of   energy $E$   for  a  
soliton 
   using the Gaussian ansatz
$\phi({\bf r})=\exp(-x^2/2w_x^2-y^2/2w_y^2-z^2/2w_z^2)/(\sqrt{w_xw_yw_z}\pi^{3/4})$, with $E_{\mathrm{kin}}=1/4w_x^2+1/4w_y^2+1/4w_z^2, E_{\mathrm{trap}}
=-V_y \exp(-w_y^2)$,  $E_{\mathrm{int}} 
=[a+a_{dd} s(k_x,k_y)-a_{dd}]/[{\sqrt{2\pi}w_xw_yw_z}]$ and
\begin{eqnarray}
\label{energy}
s(k_x,k_y)= \int_0^1 \frac{3\kappa_x\kappa_y u^2 du }{\sqrt{1+(\kappa_x^2-1)u^2}  \sqrt{1+(\kappa_y^2-1)u^2} },
\end{eqnarray}
where $\kappa_x=w_x/w_z, \kappa_y=w_y/w_z$. 


We perform    numerical simulation of the 3D GP equation (\ref{gp3d})
using   the split-step  Crank-Nicolson 
method  \cite{Muruganandam2009}.  
The dipolar term is
treated by  fast Fourier transformation  \cite{jb}.
The   error of the reported numerical results 
is less than 1 $\%$.    We present in figure \ref{fig1}
(a)  the 3D contour   of the  axially-symmetric  
bright soliton  for     $g=50$, $g_{dd}=-15$ and $V^{1D}_{\mbox{OL}}=-2\cos(2z)$.
In figure \ref{fig1} (b), we show the    anisotropic bright soliton for 
   $g=50$, $g_{dd}=20$ and $V^{1D}_{\mbox{OL}}=-2\cos(2y)$. For the
anisotropic soliton, 
the numerical energy is $-0.752$ in agreement with the energy $-0.739$
obtained from the 
minimization in (\ref{energy}). 
The anisotropy in the 
$x-z$ plane in figure \ref{fig1} (b)  is due to dipolar interaction.
In figure \ref{fig1} (c) we show an axially-symmetric 
vortex soliton 
on 1D OL, $V^{1D}_{\mbox{OL}}=-2\cos(2z)$ for $g=5$ and $g_{dd}
=-9$. A relatively large $|g_{dd}|$  is needed to 
overcome the centrifugal barrier and stabilize a vortex soliton.  
The bright solitons of {figures} \ref{fig1} (a) and (b) are stable 
in real-time propagation. 
However,  the vortex soliton {with the parameters of figure 
\ref{fig1} (c)}
suffers from 
transverse instability at large times ($t>70$), which eventually leads to its destruction \cite{transverse}.   The snapshot 
of the vortex soliton 
after real-time 
propagation at $t=70$ in figure 
\ref{fig1} (d) does not, however,  show any distortion or sign of instability. 
 The 
numerical 
energy 
and root-mean-square (rms) sizes   of the  
solitons of
 {figures} \ref{fig1}    are shown in Table 
I
with  variational results in the axially-symmetric case. 
{ In this table we also show the parameters $a_{dd}$, $\alpha$, and $N$ 
for these solitons for $^{52}$Cr and $^{164}$Dy stabilized by a laser of
 wavelength $\lambda=10,000$ \AA,   
 and atomic scattering length $a=5a_0$ obtained 
using a Feshbach resonance.   }

\begin{table}
\label{I}
\caption{Numerical ($n$) and variational ($v$) energy and 
rms sizes   $E,\langle x\rangle, \langle y\rangle, \langle z\rangle$
of  bright (br) and vortex (vor) solitons. { Experimental parameters 
$a_{dd}$ ($15a_0$ for $^{52}$Cr and 130$a_0$ for $^{164}$Dy), $\alpha$, $N$ for realizing these solitons
are given for scattering length $a=5a_0$ and  wavelength $\lambda=10000$ \AA.   } }
 
\centering
\label{table:1}
\begin{tabular}{lrrcccccccc}
\hline
&$g$ & $g_{dd}$  &$a_{dd} $ $ (a_0)$&$\alpha$ & $N$   & $E$  &  $\langle x\rangle$  & $\langle y\rangle$  &
  $\langle z\rangle$  \\
\hline
$n$, br& 50 &  $-15$ &15  &$-0.418$ &2400 & $-0.836$ & 2.00 &  2.00 & 0.450  \\
$v$, br&   50 & $-15$ & 130&$-0.0482$ &2400 &$-0.814$ & 2.086 & 2.086 & 0.423 \\
$n$, br&50 &  $20$  & 15 &  0.557 &2400 &  $-0.752$ & 0.533&  1.38 & 3.65  \\ 
$n$, vor& 5& $-9$ & 130  &$-0.289$ &240 & $-0.744$   & 4.02  &4.02  &0.663                    \\
\hline
\end{tabular}
\end{table}

\begin{figure}[!t]
\begin{center}
\includegraphics[width=\linewidth, clip]{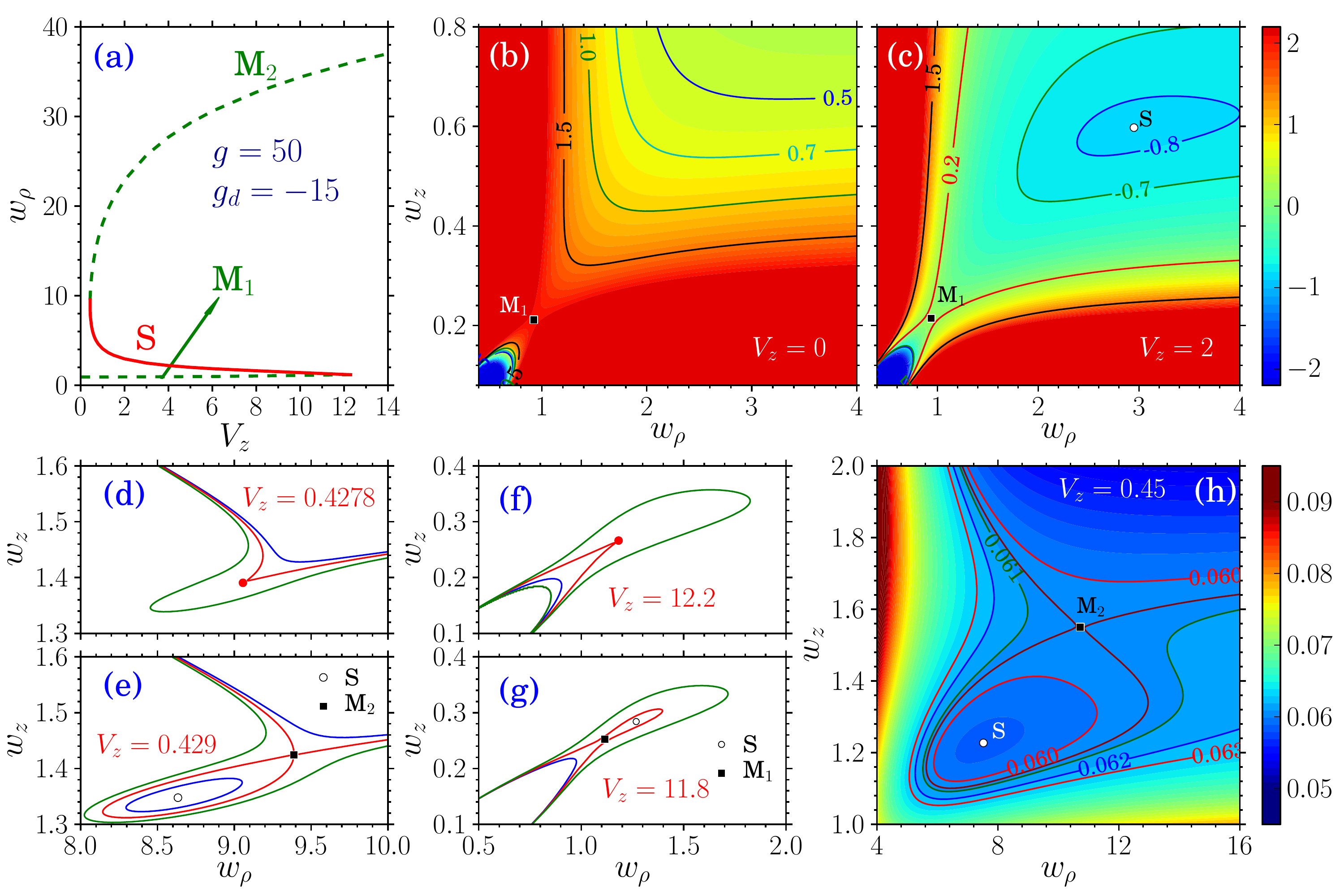}
\end{center}

\caption{(Color online) (a) Bifurcation diagram  showing saddle-center bifurcation
involving  the stable (center, $S$) and {unstable stationary} 
states (saddles, $M_1$ and $M_2$)
 in the $w_\rho$ versus $V_z$ plot. 
{The lines correspond to solutions of the variational equations. The thresholds for 
saddle-center bifurcation are at $V_z=0.4278$ and 12.2.}  
(b) $-$ (h) 
The equal-energy contours 
in the 
 $w_\rho$ versus $w_z$ phase plot 
for different $V_z$. { A center   (open circle) $S$
corresponds to a minimum of energy surrounded by closed loops
in these contour plots. A saddle (solid square), $M_1$ or $M_2$, corresponds to intersection of two equal-energy lines.
In (d) $-$ (g) the appearance of a saddle  and a
center  
for a small change of $V_z$ near the thresholds at $V_z=0.4278$ and 12.2
is illustrated.}
In all cases 
  $g=50$ and $g_{dd}=-15.$
}

\label{fig2}
\end{figure}

For the axially-symmetric bright soliton,
we have a conservative system with two degrees of freedom $w_\rho$ and $w_z$
with  Lagrangian (\ref{lag}). The stable state 
appears and disappears by saddle-center bifurcation \cite{scb} as  
  $V_z$ is increased as shown in figure \ref{fig2} (a)  for $g=50$ and $g_{dd}=-15$, where the 
  {unstable stationary} ($M_1$ and $M_2$) and stable ($S$) states are shown 
in the $w_\rho$ versus $V_z$ plot. For small $V_z (<0.4278)$ there  exists 
only the {unstable stationary} state $M_1$. At $V_z=0.4278$ and $w_\rho \approx 10$ a stable ($S$) and a  {unstable stationary} ($M_2$) state 
appear ``out of nothing"     by saddle-center bifurcation. With 
further increase of $V_z$,  the center $S$
 comes towards the saddle $M_1$
and the two disappear  ``to nothing" by a reverse (sub-critical) saddle-center bifurcation at $V_z=12.2$, whereas the state $M_2$ moves towards infinity \cite{scb}. 
 We show the equal-energy variational 
contours   in the $w_z$ versus $w_\rho$ phase plot 
for different $V_z$ in {figures} \ref{fig2} (b) $-$ (h), where the 
positions of the stable and {unstable stationary} states are also shown. Figures \ref{fig2}  (d) $-$
(g) show  close ups of the appearance and disappearance of 
the  state S by   saddle-center bifurcations.

To perform a linear stability analysis of the axially-symmetric
 states,
 we rewrite 
 (\ref{f3}) and (\ref{f4})     as \cite{stability}
\begin{eqnarray}
\dot x_1 & = x_3,\;\;\;
\dot x_2 = x_4, \\
\dot x_3&= \frac{1}{x_1^3}+\frac{1}{\sqrt{2 \pi}}\frac{N}{x_1^3x_2}\left[2a-a_{dd}e\left(\frac{x_1}{x_2}\right)
\right],\\
\dot x_4&= \frac{1}{x_2^3}-\frac{4V_zx_2}{\exp(x_2^2)}+\frac{1}{\sqrt{2 \pi}}\frac{2N}{x_1^2x_2^2}\left[a-a_{dd}h\left(\frac{x_1}{x_2}\right)\right],
\end{eqnarray}
where $(x_1, x_2, x_3, x_4) \equiv  (w_\rho, w_z, \dot w_\rho, \dot w_z)$.
These   equations for   widths can be written as $\dot {\mathbf x}\equiv {\mathbf f}({\mathbf x})$, 
where ${\mathbf x} \in \{x_1, x_2, x_3, x_4\}$. If ${\mathbf x}^{(0)}$ denote the fixed points with $\dot {\mathbf x}=0$, so that 
${\mathbf f}({\mathbf x}^{(0)})=0$, then the linearization matrix  is $J \equiv \partial {\mathbf f}({\mathbf x})/
\partial {\mathbf x}|_{{\mathbf x}={\mathbf x}^{(0)}}$ \cite{stability}. An examination of   eigenvalues  of $J$  reveals 
the nature of stability of the states.  
The eigenvalues come in pairs $\pm \lambda$ and lead to exponential growth unless all  of them  
are  imaginary corresponding to a spectrally stable equilibrium, which is of interest in the 
present context.
For the parameters of figure \ref{fig1} (a),  
there exist the saddle $M_1$
at ${\mathbf x}^{(0)}\equiv(0.9379, 0.2147, 0, 0)$, and the center $S$ at  ${\mathbf x}^{(0)}\equiv (2.9497, 0.5976, 0, 0)$, and the saddle $M_2$ at ${\mathbf x}^{(0)}\equiv (22.8943,
2.3485, 0, 0)$ of which $S$ and $M_2$  are shown in figure \ref{fig2} (h). The eigenvalues are ($\pm 4.8644$, $\pm 19.9241i$)
for  $M_1$, 
($\pm 4.4835i$, $\pm 0.2274i$) for 
  $S$, and ($\pm 0.4751$, $\pm 0.0018i$) for $M_2$. The center with pairs of pure {\it imaginary}
eigenvalues confirm its stability.

\begin{figure}[!b]
\begin{center}
\includegraphics[width=\linewidth,clip]{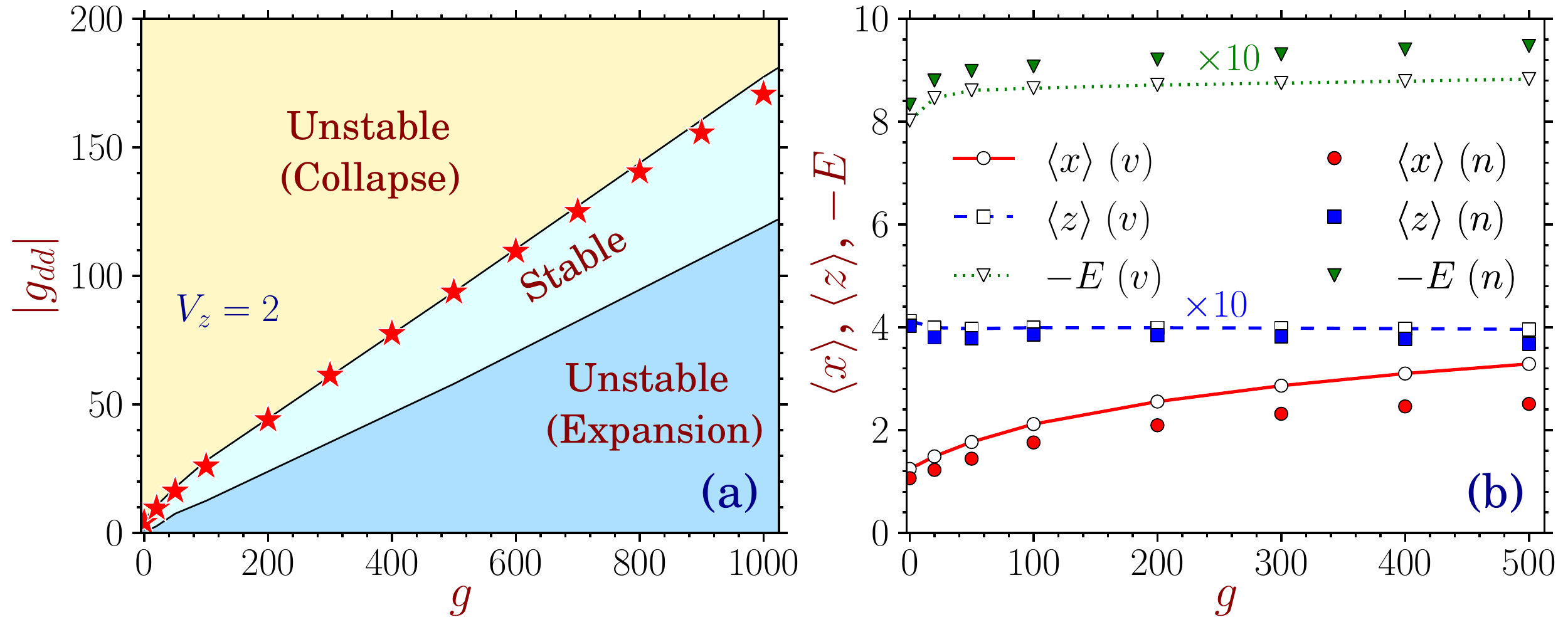}
\end{center}

\caption{(Color online) (a) The  phase plot of $|g_{dd}|$ versus $g$ from 
variational analysis of the axially-symmetric solitons
showing  the region of stable solitons for $V_z=2$. The $\star$'s denote 
the numerical points showing the stable-{unstable} boundary.
(b)  The {numerical (n) and variational (v)} rms sizes and energy versus $g$ for
$|g_{dd}|$ corresponding to the $\star$'s in (a). 
}
\label{fig3}
\end{figure}

Using variational equations, 
we analyze the appearance of axially-symmetric  bright 
solitons  
 using the phase plots of $|g_{dd}|$ versus $g$ for   $V_z=2$ 
in figure \ref{fig3} (a).  
For $|g_{dd}|$ in a window of
critical values, stable bright solitons can be formed. For smaller $|g_{dd}|$,
there is too much repulsion and the system expands to infinity 
and for 
larger $|g_{dd}|$, there is too much attraction leading to collapse
allowing only {unstable stationary} states. In figure \ref{fig3} (b) we plot the variational 
and numerical rms sizes and energies for $|g_{dd}|$ corresponding to the numerical points ($\star$) 
in figure \ref{fig3} (a).

\begin{figure}
\begin{center}
\includegraphics[width=\linewidth,clip]{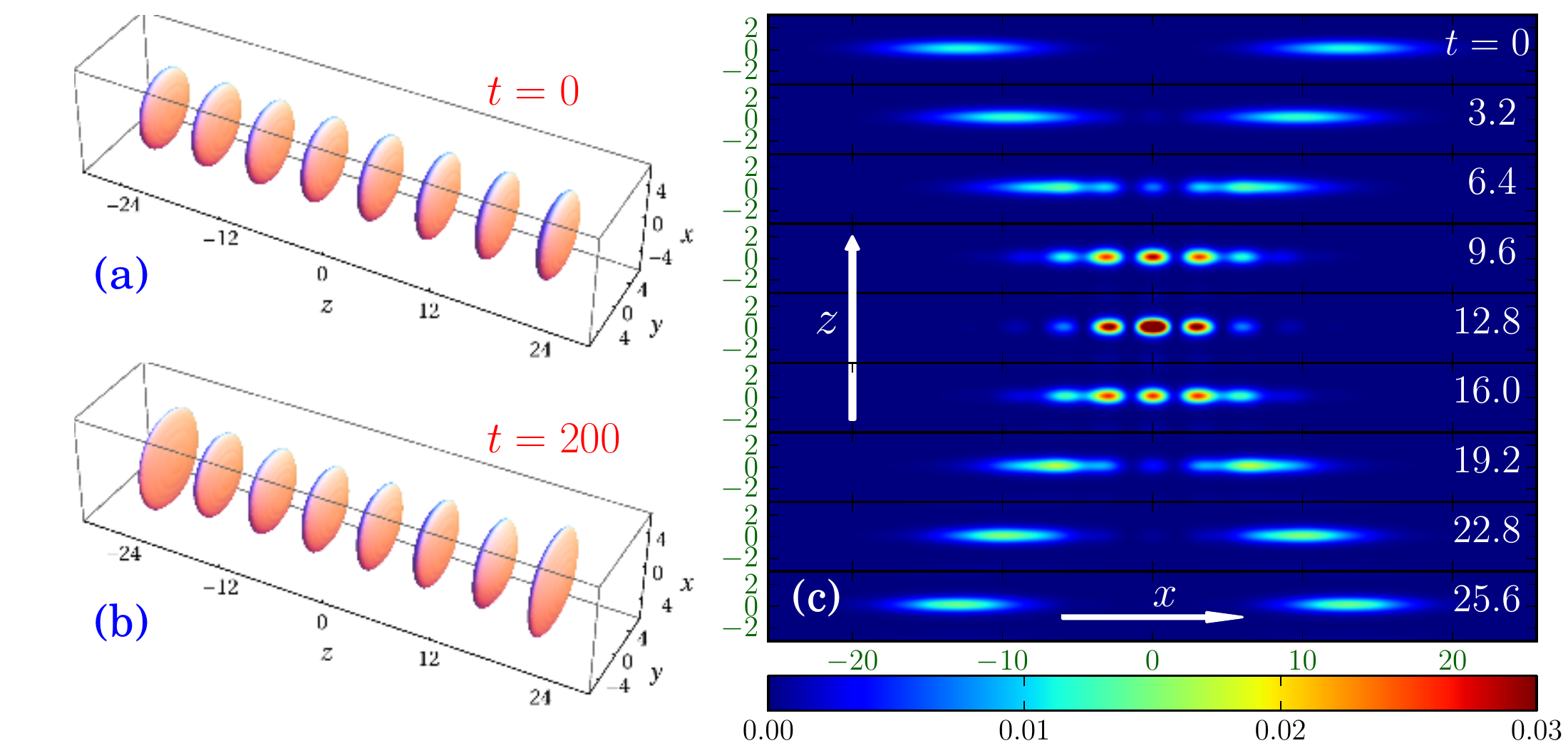}
\end{center}

\caption{(Color online)   3D contour of the stable array of eight  
axially-symmetric
solitons each with $g=-g_{dd}=5$ on   OL  
$V^{1D}_{\mbox{OL}}=
-2\cos(2z)$ at (a) $t=0$ and at (b) $t=200$.
The density at contour is 0.001. (c) Contour plot of density 
$|\phi(x,y,0,t)|^2$ of two colliding solitons each with 
$g=5$ and $g_{dd}=-3$  and  velocity 1, before, during, and after 
collision.}
\label{fig4}
\end{figure}

Next, we consider  a stable 
soliton array 
by mounting   tiny DBEC bright solitons 
along the supporting OL sites. Such solitons will be interacting due to long-range 
dipolar interaction. We prepare a stable array of axially-symmetric 
solitons  each  with $g=-g_{dd}=5$
by putting them 
on alternate sites
of OL $V^{1D}_{\mbox{OL}}=-2\cos(2z)$ 
at $x=y=0, z=\pm (2n+1)\pi, n=0,1,2,3$. Such an array of bright solitons 
with small $g$ and $|g_{dd}|$ are stable, whereas those of large $g$ and $|g_{dd}|$, e.g. the one
of figure \ref{fig1} (a),  are unstable.
 In figure \ref{fig4} (a) and (b) we show the 
initial array and the  
the final profile after real-time 
propagation at $t=200$. A similar array of anisotropic solitons has a finite life 
and is destroyed at large times.    

Finally,  we investigate the collision of two axially-symmetric
bright solitons with 
$g=5$ and $g_{dd}=-3$ each, placed at 
$x=\pm  12.8$, $y=z=0$ at   $t=0$. Each soliton is given a velocity of 
$v=1$ towards center $x=0$ by a phase factor $\exp(\pm ix)$,
respectively,
in the initial wave functions. The collision dynamics is illustrated in figure 
\ref{fig4} (c) where we show the snapshots of contour plots of density 
$|\phi(x,0,z,t)|^2$ at different times. The solitons come towards 
each other, interact at $x=0$ and $t\approx 12.8$ and come out without  deformation 
showing their robustness. In this simulation, not only 
the parameters of two individual solitons should lead to a stable state,  
the combined nonlinearities $2g$ and $2g_{dd}$
should also correspond to a stable configuration in figure \ref{fig3} (a)
to avoid collapse during collision. { It was demonstrated in   \cite{ps} that, 
under harmonic confinement, after collision at very low velocities, 
two quasi-2D dipolar BEC solitons may merge together to form a single soliton molecule.
However, in the present case the solitons appear in a narrow window of nonlinearities $g$ 
and $g_{dd}$, as can be seen from figure \ref{fig3}.  If two equal 2D solitons, as in figure \ref{fig4} (c),  coalesce at low 
velocities, the  nonlinearities  of the merged soliton molecule will be outside the domain 
of stability in  figure \ref{fig3}. Hence, the formation of soliton molecule is mostly not possible in 
the present case. At large velocities the solitons of \cite{ps} undergo quasi-elastic collision
quite similar to the present collision shown in figure 
\ref{fig4} (c). Also, both quasi-elastic collision at large velocities and merging at low velocities 
of two quasi-1D solitons under transverse harmonic confinement was illustrated in \cite{adhisol}.
}

To summarize,
 we studied different types of 2D bright 
 solitons in a DBEC with repulsive atomic interaction using the solution of the 3D GP equation.
 Anisotropic stable 2D 
bright solitons  in DBEC are possible 
 on a weak 1D OL perpendicular to the polarization direction.
Axially-symmetric stable 2D
bright and vortex solitons in DBEC can be generated 
 on  a weak 1D OL along the polarization direction 
when the dipolar interaction 
is tuned to negative values \cite{condip}. 
{In this sign-changed dipolar-interaction configuration, bright and vortex solitons 
 are stable due to the long-range attractive dipolar interaction in the quasi-2D shape.}
In the axially-symmetric case, an 1D stable array of tiny solitons, 
placed on alternate OL sites,
can be made. Such 1D array with empty OL sites between solitons  
is of interest in condensed matter physics
\cite{band,lewen} and bears some similarity with stable checkerboard pattern of 
DBEC on 2D OL with empty sites in between
\cite{barbara}, both  arising due to dipolar interaction. The elastic 
nature of collision of two axially-symmetric solitons is also demonstrated.   
With present technology these stable 2D solitons and their 1D arrays 
can be created  and studied  in laboratory.
 
\ack

We thank {Prof. B. A. Malomed for a critical reading of the 
manuscript and}
FAPESP (Brazil),   CNPq (Brazil),    DST (India),   and CSIR  (India)
for partial support.

\end{document}